\documentstyle[multicol,aps,epsfig,prb]{revtex}

\begin{document}
\draft
\title{Tunneling of correlated electrons in ultra high magnetic field}
\author{Shan-Wen Tsai$^1$, D. L. Maslov$^1$ and L. I. Glazman$^2$}
\address{$^{1)}$Institute for Fundamental Theory and \\
Department of Physics, University of Florida, Gainesville,
FL 32611\\
$^{2)}$Theoretical Physics Institute, University of Minnesota, Minneapolis, 
MN 55455\\
}
\date{\today}
\maketitle

\begin{abstract}
Effects of the electron-electron interaction on tunneling into a metal in
ultra-high magnetic field (ultra-quantum limit) are studied. The range of
the interaction is found to have  a decisive effect both on the nature of the 
field-induced instability of the ground state and on the properties of the
system at energies above the corresponding gap. For a short-range repulsive
interaction, tunneling is dominated by the renormalization of the coupling 
constant, which leads eventually to the charge-density wave instability. For 
a long-range interaction, there exists an
intermediate energy range in which the conductance obeys a power-law scaling
form, similar to that of a 1D Luttinger liquid. The exponent is
magnetic-field dependent, and more surprisingly, may be positive or
negative, {\it i. e.}, interactions may either suppress or enhance the
tunneling conductance compared to its non-interacting value. At energies near
the gap, scaling breaks
down and tunneling is again dominated by the instability, which in this
case is an (anisotropic) Wigner crystal instability.
\end{abstract}


\begin{multicols}{2}
Low-dimensional systems exhibit ``zero-bias anomalies'' in tunneling
(non-linearities of the current-voltage characteristics at small
biases), which reflect the renormalization of the density of states by
the electron-electron interaction. In particular, tunneling into a
one-dimensional (1D) metal (Luttinger liquid) is characterized by a
power-law suppression of the tunneling conductance, $g_{T}$.  A
three-dimensional metal placed in a strong magnetic field that
depopulates all but one Landau levels (ultra-quantum limit, UQL)
provides an example of a very special quasi-1D system. It is
well-known \cite{Brazovskii,Fukuyama,Yakovenko} that repulsive
interactions in the UQL lead to a charge-density wave (CDW) or a
Wigner crystal instability of the ground state. This has been
confirmed, for example, by experiments on graphite in high magnetic
fields\cite{cdw}. The most complete analysis of this instability for
the case of a short-range interaction was performed in
Ref. [\onlinecite{Yakovenko} ] by solving the renormalization-group
(RG) equation for the interaction vertex. On the other hand, it has
recently been shown that for the case of a long-range (Coulomb)
interaction $g_{T}$ exhibits a power-law, Luttinger-liquid-like
behavior at sufficiently high energies\cite{Biagini}. This result was
obtained in two ways -- via a perturbative Hartree-Fock procedure,
which results in the RG equation for the transmission amplitude, and
via bosonization in the coherent state basis -- neither of which took
into account the renormalization of the interaction vertex. In this
paper, we combine the approaches of
Refs.~[\onlinecite{Brazovskii,Yakovenko} ] and [\onlinecite{Biagini} ]
to study the behavior of the tunneling conductance in the whole energy
interval from the Fermi energy down to the 
energy associated with the instability. This is
accomplished by solving a coupled system of the RG equations for both
the interaction vertex and transmission amplitude. We find that for a
long-range interaction there is a parametrically wide energy interval
in which the flow of the transmission amplitude already results in its
power-law scaling with energy, whereas the renormalization of the
vertex is not yet important. In this interval, the system behaves as a
Luttinger liquid and the results of Ref. [\onlinecite{Biagini} ] are
applicable. At energies close to the gap, the power-law scaling breaks
down and tunneling is dominated by the  electron properties in
the vicinity of the critical point. For a short-range interaction,
the critical point dominates tunneling at all energies, and there is
no Luttinger-liquid behavior. We also find that above a certain
magnetic field,
when the lowest Landau level is strongly depopulated, the interaction {\em %
enhances }the tunneling conductance of the Luttinger-liquid regime above its
free-electron value. Such an unusual behavior receives a natural explanation
in terms of electron scattering from the Friedel oscillation near the
tunneling barrier. 

First we review the
procedure of finding the effect of the electron-electron interaction on the
tunneling conductance of a metal in the UQL\cite{Biagini}. Let the magnetic
field be perpendicular to the contact plane ($z=0$) that separates two
metallic sides. We consider both the symmetric configuration, where both
sides are in the UQL, and the asymmetric one, where one of the sides is made
of a high-carrier concentration and/or dirty metal so that, in the first
approximation, the magnetic field does not affect that side. The
transmission and reflection amplitudes for non-interacting electrons, $t_{0}$
and $r_{0}$, are assumed to be known. We choose the Landau basis for the
free electron wavefunctions (non-interacting electrons in the UQL and 
in the presence of the barrier): 
\[
\psi _{p_{z},p_{x}}^{(0)}({\bf r})=\psi _{p_{z}}(z)\chi _{p_{x}}(x,y), 
\]
where 
\begin{eqnarray}
\psi _{p_{z}}(z) & = & \theta(z) t_{0}e^{ip_{z}z}  
+ \theta(-z) (e^{ip_{z}z}+r_{0}e^{-ip_{z}z}) \nonumber \\ 
\psi _{-p_{z}}(z) & = & \theta(z) (e^{-ip_{z}z}+r_{0}e^{ip_{z}z})  
+ \theta(-z) t_{0}e^{-ip_{z}z}. 
\end{eqnarray}
(Here and thereafter, we set $\hbar =1$ and also define the magnetic length
as our unit of length: $\sqrt{\hbar c/eB}\equiv 1$).

Backscattering of magnetically quantized electrons at the barrier gives rise
to a Friedel oscillation in the charge density, whose amplitude decays away from the barrier as $%
z^{-1}$. Correspondingly, the exchange, $V_{ex}({\bf r},{\bf r}^{\prime })$,
and Hartree, $V_{H}({\bf r})$, potentials calculated using exact (in the
presence of the barrier) but otherwise free wavefunctions, also exhibit
rapid $2k_{F}$-oscillations and decay as $z^{-1}$ away from the barrier. The
first-order correction to the wavefunction due to interaction is given by 
\begin{eqnarray}
\delta \Psi _{p_{z},p_{x}}({\bf r}) &=&\int d{\bf r}^{\prime }\int d{\bf r}%
^{\prime \prime }G({\bf r},{\bf r}^{\prime };E)\left[V_{ex}({\bf r}^{\prime },{\bf %
r}^{\prime \prime }) +  \right. \nonumber \\
&&\left. \delta ({\bf r}^{\prime }-{\bf r}^{\prime \prime })V_{H}({\bf r}^{\prime
\prime })\right]\Psi _{p_{z},p_{x}}^{(0)}({\bf r}^{\prime \prime }),  \label{dpsi}
\end{eqnarray}
where $G({\bf r},{\bf r}^{\prime };E)$ is the Green's function in the
presence of the magnetic field. When the momentum transfer ($2p_{z}$)
matches the wavevector of the Friedel oscillation ($2k_{F}$), a slow ($z^{-1}
$) decay of the oscillations in $V_{ex}$ and $V_{H}$ leads to a logarithmic
singularity in $\delta \Psi ({\bf r})$, and consequently, in $\delta t$. For
electron's energy close to the Fermi energy, we have 
\begin{equation}
t=t_{0}\left[ 1-|r_{0}|^{2}c\Gamma _{0}\ln \left( {W/E}\right) \right] 
\label{1stlog}
\end{equation}
where $c=1(1/2)$ for symmetric (asymmetric) configuration, $W\sim E_{F}$ is
the effective bandwidth, $\Gamma _{0}\equiv\Gamma ({\bf q}_{\perp }=0)$, and $%
\Gamma ({\bf q}_{\perp })$ is expressed via the Fourier transform of the
interaction potential $U_{0}(q_{z},{\bf q}_{\perp })$: 
\begin{eqnarray}
\Gamma ({\bf q}_{\perp })=\frac{1}{(2\pi )^{2}v_{F}}\Bigl[ &&\int \frac{%
d^{2}k_{\perp }}{2\pi }e^{i{\bf q}_{\perp }\cdot {\bf k}_{\perp }-k_{\perp
}^{2}/2}U_{0}(0,{\bf k}_{\perp })  \nonumber \\
&&-U_{0}(2k_{F},{\bf q}_{\perp })e^{-q_{\perp }^{2}/2}\Bigr].  \label{gamma0}
\end{eqnarray}
The interaction potential, $U_{0}$,  can be both of
long- and short-range. The first situation corresponds to a single-band
metal, in which there are no more charge carriers besides those already in
the UQL. In this case, $U_{0}(q_{z},{\bf q}_{\perp })=4\pi
e^{2}/(q^{2}+\kappa ^{2})$  (for $q\ll k_{F}),$ where $\kappa =\omega
_{p}/v_{F}$ and $\omega _{p}$ is the plasma frequency at $B=0.$ The
perturbation theory works if $e^{2}/v_{F}\simeq \kappa ^{2}\ll 1$ , which
means that the interaction is long range. As the magnetic field reduces the
phase-space available for the motion along the field, $k_{F}$ and $v_{F}$ decrease
with the field (in the UQL, $k_{F},v_{F}\propto B^{-1}).$ Therefore, the
perturbation theory breaks down at sufficiently strong field so that $\kappa
\simeq 1$. The case of a short-range interaction may correspond to a
situation when a small pocket of the Fermi surface is in the UQL, whereas
other parts of the Fermi surface still contain many Landau levels. Screening
of the interaction among the magnetically quantized carriers is then mostly
due to the ``external'' carriers, and the interaction may be of short-range.

The first and second terms in Eq.(\ref{gamma0}) are the exchange and Hartree
contributions, respectively. Due to the Pauli principle, they enter Eq.(\ref{gamma0})
with opposite signs. Note that exchange
contribution involves integration over the transverse momentum ${\bf k}%
_{\perp }$, whereas the Hartree contribution enters Eq.(\ref{gamma0}) at $%
q_{\perp }=0$. This is due to a simple fact that the Friedel oscillation in
charge density, and hence in the Hartree potential, is 
translationally-invariant along the barrier, whereas the exchange potential
involves the density matrix and is thus non-local. If the interaction potential
is peaked strongly at $q=0$ ({\em e.g.}, Coulomb potential) the Hartree
contribution is enhanced due to $q_{\perp }=0$ and may in fact {\em dominate 
}over the exchange one. In this case $\Gamma _{0}$ in Eq.(\ref{1stlog}) is
negative and thus the transmisson amplitude is {\em enhanced }by the
interaction. Such an unusual behavior should be contrasted, {\em e.g.}, to a
strictly 1D case \cite{Yue}, in which $\Gamma _{0}=U_{0}(0)-U_{0}(2k_{F}) $
and is thus positive for any ``realistic'' interaction potential, {\em i.e., 
}such that $U_{0}(q)$ is a monotonically decreasing function of $q$. In our
case, due to the 3D nature of the problem, the exhange and Hartree 
contributions
are not simply given by the corresponding forward and backscattering
amplitudes but also involve averaging over the transverse direction. As a
result, $\Gamma _{0}$ can take either sign.

At the next order, $V_{ex}$ and $V_{H}$ are re-calculated using the
corrected wavefunctions and are substituted back into Eq.(\ref{dpsi}). The
higher order corrections to $\delta \Psi ({\bf r})$ are higher powers of
logs. Similarly to tunneling through a barrier of weakly
interacting electrons in 1D,\cite{Yue} summation of the most divergent
corrections to $t$ in all orders of the perturbation theory can be performed
via an RG equation 
\begin{equation}
\frac{dt}{d\xi }=-c\Gamma _{0}\ t\ (1-|t|^{2}),
\label{rgtonly}
\end{equation}
where $\xi \equiv \ln \left( {W/E}\right) $ and $t(\xi =0)=t_{0}$. The solution of (\ref{rgtonly}) is 
\begin{equation}
t={\frac{t_{0}(E/W)^{c\Gamma _{0}}}{\sqrt{|r_{0}|^{2}+|t_{0}|^{2}(E/W)^{c%
\Gamma _{0}}}}}.  \label{tfull}
\end{equation}
For small values of 
\mbox{$\vert$}%
$t_{0}|$ (high barrier), Eq.(\ref{tfull}) reduces to a power-law form 
$
t=t_{0}\left( E/W\right) ^{c\Gamma _{0}} 
$.
The tunneling conductance is related to the transmission amplitude via the
Landauer formula.

The power-law solution for $t$ is obtained using the bare interaction
vertex. However, the vertex is also subject to renormalization. The flow of $%
\Gamma ({\bf q}_{\perp },\xi )$ is described by the integro-differential RG
equation \cite{Yakovenko}, which has a more transparent meaning when written
for the Fourier transform of $\gamma ({\bf r}_{\perp },\xi )=1/(2\pi)^2\int
d^{2}q_{\perp }\Gamma ({\bf q}_{\perp },\xi )e^{-i{\bf q}_{\perp }\cdot {\bf %
r}_{\perp }}$: 
\begin{eqnarray}
&&\frac{d\gamma ({\bf r}_{\perp },\xi )}{d\xi }=\int 
dr_{\perp }^{\prime }\gamma ({\bf r}_{\perp }^{\prime },\xi )\gamma (%
{\bf r}_{\perp }-{\bf r}_{\perp }^{\prime },\xi )  \nonumber \\
&&\times \left( 1-e^{i{\bf r}_{\perp }\wedge {\bf r}_{\perp }^{\prime
}}\right) ,
\label{rggamma}
\end{eqnarray}
where {\bf a}$\wedge {\bf b}=a_{x}b_{y}-a_{y}b_{x}$.
The initial condition for Eq. (\ref{rggamma}) is given by $\Gamma ({\bf q}%
_{\perp },\xi =0)=\Gamma ({\bf q}_{\perp })$, where $\Gamma ({\bf q}_{\perp
})$ is defined by Eq. (\ref{gamma0}). The flow of the vertex affects the RG
equation for $t$, in which now the bare vertex has to be replaced by the
renormalized one: 
\begin{equation}
\frac{dt}{d\xi }=-c\Gamma ({\bf q}_{\perp }=0,\xi )\ t\ (1-|t|^{2})\ . 
\label{rgt}
\end{equation}
Eqs. (\ref{gamma0}), (\ref{rggamma}) and (\ref{rgt}) provide a full
description of tunneling into a three-dimensional metal in the UQL.

Few words about the general features of Eq.(\ref{rggamma}) are now in place.
Physically, $\gamma ({\bf r}_{\perp },\xi )$ plays the role of an
interaction potential between two electrons whose guiding centers are 
separated by a ``distance'' 
\mbox{$\vert$}%
${\bf r}_{\perp }|$ across the magnetic field. (Note
however that $\gamma ({\bf r}_{\perp },0)$ is related to the Fourier
transform of the bare interaction, cf. Eq. (\ref{gamma0}).) The first and
second terms in the brackets in Eq. (\ref{rggamma}) correspond to the Peierls
and Cooper interaction channels, respectively. A strictly 1D case is
recovered by putting all guiding centers onto the same line: ${\bf r}_{\perp }=%
{\bf r}_{\perp }^{\prime }.$ In this case, ${\bf r}_{\perp }\wedge {\bf r}%
_{\perp }^{\prime }=0$ and the Peierls and Cooper channels cancel each other
exactly. This cancellation reflects the well-known fact that there are no
instabilties for 1D (spinless) electrons and the system remains in the
massless phase (Luttinger liquid) down to the lowest energies. In the UQL
case ${\bf r}_{\perp }\neq {\bf r}_{\perp }^{\prime }$, Peierls and Cooper
channels do not cancel each other, and an instability becomes possible.

For a short-range interaction, the $q_{\perp }$-dependence of $U_{0}$ can be
neglected: $U_{0}(0,{\bf q}_{\perp })\rightarrow u_{0}$, $U_{0}(2k_{F},{\bf q%
}_{\perp })\rightarrow u_{2k_{F}}$ and $\Gamma ({\bf q}_{\perp })=\left(
u_{0}-u_{2k_{F}}\right) /(2\pi )^{2}v_{F}e^{-q_{\perp }^{2}/2}$. For
repulsion, the solution of Eq. (\ref{rggamma}) with the Peierls term alone
is
\cite{Brazovskii,Yakovenko} 
\begin{equation}
\Gamma ({\bf q}_{\perp },\xi )=\left( \Gamma ^{-1}({\bf q}_{\perp })-\xi
\right) ^{-1}.  \label{pole}
\end{equation}
The vertex diverges most rapidly at ${\bf q}_{\perp }=0$ and the 
equation for the pole $\xi =\Gamma ^{-1}(0)$
yields the gap $\Delta =W\exp (-1/\Gamma
_{0})$. As the electron density remains homogeneous across the field, the
instability correspond to a CDW along the field. Inclusion of the Cooper
term merely changes the magnitude of the gap \cite{Yakovenko}. The RG
equation for $t(\xi )$ [Eq.(\ref{rgt})] with $\Gamma ({\bf {q}_{\perp },\xi )%
}$ from Eq.(\ref{pole}) allows for an analytic solution: 
\begin{equation}
t=\frac{t_{0}f(E)}{\sqrt{|r_{0}|^{2}+|t_{0}|^{2}f^{2}(E)}}\ ,  \label{anst}
\end{equation}
where $f(E)\equiv \lbrack \Gamma _{0}\theta (E-\Delta )\ln E/\Delta ]^{c}$.
Near the gap ($E\approx \Delta $), the tunneling conductance $g_{T}\propto
|t|^{2}$ assumes a {\em universal} form 
\begin{equation}
g_{T}\propto (E-\Delta )^{2c}\theta (E-\Delta ).  \label{neargap}
\end{equation}
Comparing $t$ from Eq.(\ref{anst}) with Eq.~(\ref{tfull}), obtained without taking into
account the flow of the interaction vertex $\Gamma$, we see that
the two coincide only at the lowest order in the interaction (first
log-correction to $t_{0}).$ In contrast to Eq.(\ref{tfull}), the full
solution shows that the higher order logarithmic terms do not sum up into a
power-law. This is already seen from the fact that the first 
log-correction to $t_{0}$ [Eq.(\ref{1stlog})] becomes of order $t_{0}$ at $%
E\simeq $ $\Delta ,${\em \ i.e.}, when the renormalization of 
$\Gamma$ cannot be neglected. In other words, the CDW critical point dominates the
physical properties in the entire energy interval from $W$ to $\Delta .$

We now turn to the case of a long-range interaction. It is instructive to
consider first an example of an ultra-long interaction: $U_{0}({\bf r}%
)=const\rightarrow \gamma ({\bf r}_{\perp },0)\propto \delta ({\bf r}_{\perp
}).$ To analyze the flow of $\gamma ({\bf r}_{\perp },\xi )$ for small $\xi $
it suffices to substitute $\gamma ({\bf r}_{\perp },0)$ into
the right-hand side of Eq.(\ref{rggamma}), upon which the latter vanishes.
It means that the ultra-long range interaction is equivalent to a strictly
1D case, when $\gamma ({\bf r}_{\perp },\xi )$ is invariant under
renormalization. The example just given is certainly not 
realistic:
even for the Coulomb interaction 
the ground state is unstable with regard to the CDW formation
\cite{Fukuyama}. However, remnants of
this behavior survive for the Coulomb potential in a sense that $\gamma (%
{\bf r}_{\perp },\xi )$ flows {\em slower} than $t(\xi ).$ As a result,
there exists an intermediate energy scale, $E_{PL},$ ($\Delta \ll E_{PL}\ll
W)$ such that at $E\simeq E_{PL}$ the higher order logarithmic terms
generated by Eq.(\ref{rgt}) have already summed up into a power-law form but
the renormalization of $\Gamma $ is not significant yet.  $E_{PL}$ may be
defined by the condition that the lowest-order correction to $t_0$ becomes equal to $t_0$: 
\[
\xi _{PL}=\ln \left( W/E_{PL}\right) =1/c\Gamma _{0}.
\]
For a screened Coulomb potential, Eq.(\ref{gamma0}) together with
the condition $\kappa \ll k_F \lesssim 1$ yields
\begin{equation}
\Gamma _{0}=\left( e^{2}/\pi v_{F}\right) \left[ |\ln \kappa |-\left(
4k_{F}{}^{2}\right) ^{-1}\right] .  \label{alpha}
\end{equation}
A simple estimate for  $E_{PL}$ can be  obtained if the first (exchange) term
in Eq.(\ref{alpha}) is larger than the second (Hartree) one, in which case 
$E_{PL}\simeq
W\exp (-\pi v_{F}/ce^{2}|\ln \kappa |).$ At the same time, for $k_{F}\simeq
1,$ the gap is estimated as $\Delta =W\exp (-av_{F}/e^{2}),$ where $a\simeq
1.$ We see  that $E_{PL}\gg \Delta $ due to the presence of a large
logarithm under the exponential. 

\begin{figure}
\epsfxsize=3in \epsfbox{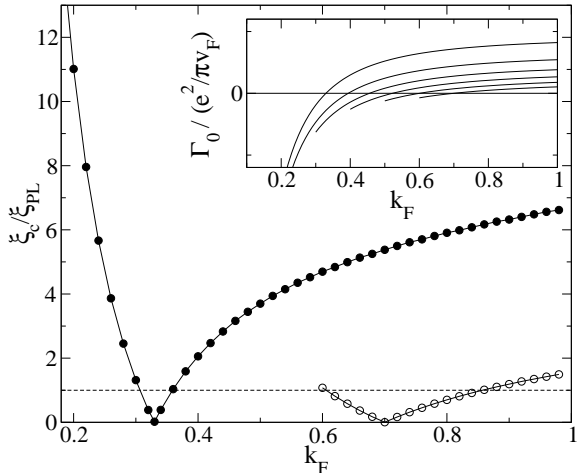}
\caption{Ratio $\protect\xi _{c}/\protect\xi _{PL}$ for 
$\protect\kappa =0.1$ (filled circles) and $\protect\kappa = 0.6$ 
(open circles). The dashed line corresponds to the value 1. Inset: 
$\Gamma_0/(e^2/\pi v_F)$ 
as function of $k_F$ for various values of $\kappa$ (from top to bottom: $\kappa = 0.1, 0.2, 0.3, 0.4, 0.5, 0.6$).}
\label{fig:ratio}
\end{figure}

As it has already been pointed out, $\Gamma _{0}$ is {\em not }%
necessarily positive: if $k_{F}$ is sufficiently small (the lowest Landau
level is strongly depleted),  the Hartree (backscattering) term dominates
and $\Gamma _{0}<0$.  The dependence of $\Gamma _{0}$ on $k_{F}$
is shown in the inset of Fig. \ref{fig:ratio} for a range of $\kappa$. Recalling that both $k_{F}$ and $\kappa $ are
magnetic-field dependent, one can express the field $B_0$ at which 
$\Gamma _{0}$
changes sign in terms of the nominal field $B_{Q}$, at which the UQL is
achieved, and the familiar gas parameter, $r_{s}$, as $B_{0} \approx 
2B_{Q}/\ln^{1/3}Cr_{s}^{-1},$ where $C\simeq 1$. For $r_{s}\simeq 1$
$B_{0}\approx 2B_{Q}$.  The 
gap as a function of $k_F$ and $\kappa$ was obtained 
by solving the RG equation
(\ref{rggamma})
numerically 
with Eq.(\ref{gamma0}) as an initial
condition. The ratio $\xi _{c }/\xi _{PL}\equiv (\ln W/\Delta )/\ln
(W/E_{PL})$ is plotted in Fig.~\ref {fig:ratio}. All numerical results
presented here are for a symmetric contact, {\it i. e.}, $c = 1$. For
a sufficiently long-range interaction (cf. filled circles for
$\kappa=0.1$ in Fig.~\ref {fig:ratio}) there is a wide range of $k_F$
(and thus of the magnetic field) in which $\xi _{c}/\xi _{PL}>1$. Once
this condition is satisfied, there exists an energy interval $\Delta
\ll E\lesssim E_{PL}$ ($\xi_{PL}\lesssim\xi\lesssim\xi_c$) in which
the system effectively behaves as a 1D Luttinger liquid, exhibiting a
characteristic power-law scaling of the tunneling conductance. (For
$E_{PL}\ll E\ll W$ the power-law reduces to the first log-correction.)
Near the gap ($E\simeq \Delta $), the power-law behavior crosses over
to the threshold behavior [Eq. (\ref{neargap})].  If the interaction
is not sufficiently long-ranged, there is no energy interval in which
$\xi _{c}/\xi _{PL}>1$ is satisfied. This is illustrated by the open
circles in Fig. \ref{fig:ratio} ($\kappa = 0.6$). (We did not plot the
points corresponding to $k_F<\kappa$, as it would have been
incosistent with our weak-coupling approximation.)
%
\begin{figure}
\epsfxsize=3in \epsfbox{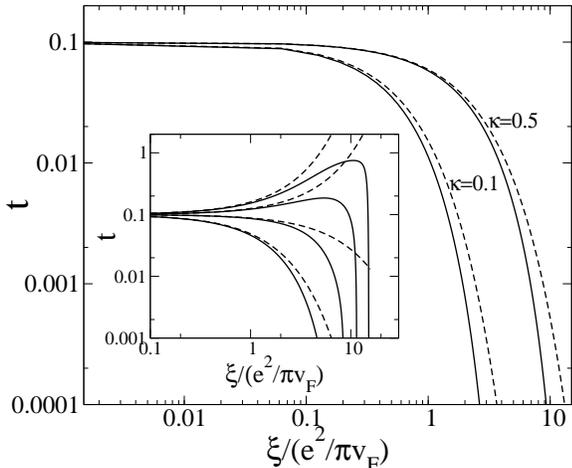}
\caption{Solid lines: solution of the full system of RG 
equations Eqs.~(8, 7, 4) for the 
screened Coulomb potential.  Dashed lines:  
solutions of Eq.~(8) with $\Gamma_0$ given by its bare value. In the 
main plot 
$t_0=0.1$, $k_F = 0.8$ and the values of $\kappa$ are indicated in the figure.
Inset: $k_F=0.4$. The curves, from bottom to top, 
correspond to $\kappa = 0.1, 0.2, 0.3$ and $0.4$.  }
\label{fig:t}
\end{figure}
A full solution of Eqs.~(\ref{gamma0},\ref{rggamma},\ref{rgt})
illustrating various regimes of the $t(E)$-dependence for 
a screened Coulomb potential is represented by solid lines 
in Fig.~\ref{fig:t} for $t_0 = 0.1$. 
Dashed lines are solutions of Eq.~(\ref{rgt}) only with $\Gamma_0$ given 
by its bare value [Eq.~\ref{gamma0}], {\it i. e.}, without taking into 
account the renormalization of the coupling constant. In the main panel,
$k_F = 0.8$. The region of small $\xi$ shows a good agreement 
between the full solution (solid lines) and the Luttinger-liquid solution 
(dashed lines). As $E$ approaches $\Delta$ ($\xi \rightarrow \xi_c$), scaling
breaks down and $t(E)$ vanishes at $E=\Delta$. In the inset 
of Fig. \ref{fig:t}, $k_F = 0.4$. The rise of $t$ with $\xi$ 
corresponds to the negative value of
the bare vertex, $\Gamma_0$. This corresponds to an enhancement of 
$g_T$ above its non-interacting value. Had the vertex 
renormalization been absent, $t$ would have reached unity in the limit 
$E\to 0$ ($\xi\to\infty$). However, the vertex renormalization 
curbs 
the rise in $t$ and eventually brings it down to zero at $E=\Delta$.
At the instability, 
the system becomes an insulator. For long-range interaction, we find that 
the vertex diverges most rapidly at finite $q_{\perp}$ 
(for $\kappa\to 0$, $q_\perp\to \approx 0.7$). This is consistent with 
previous work\cite{Fukuyama,cdw} and suggests that in this case the 
instability is of the Wigner-crystal type, with charge modulation 
both along the field and in the transverse directions.  

In conclusion, we found that the range of the electron-electron interaction
plays a crucial role both in the nature of a field-induced instability
and in tunneling at energies above the corresponding gap. For a short-range
interaction, there is a CDW instability which dominates tunneling in the whole
energy interval. For a long-range interaction, depending on the depletion 
of the lowest Landau 
level, tunneling may exhibit two types of characteristic behavior. For 
not too strong depletion, the tunneling conductance decreases as a power-law 
of the relevant energy scale at higher energies and vanishes in a 
threshold-like manner near the gap 
(which for the long-range case is the Wigner-crystal gap). 
For a stronger depletion, the 
conductance first {\it increases} as a power
of energy, then goes through a maximum and finally vanishes at the gap.
These predictions are amenable to a direct experimental verification by
tunneling into low-carrier-density materials, {\it e.g.}, graphite.

\indent We are grateful to E. V. Sukhoroukov and V. M. Yakovenko for
very instructive discussions. D. L. M. acknowledges support from the NHMFL
In-House Research Program, NSF Grant No. DMR-9703388 and Research
Corporation (RI0082). L. I. G. acknowledges support from NSF Grant No.
DMR-9731756.

\end{multicols}

\end{document}